\documentclass[reprint,superscriptaddress,amssymb,aps,twocolumn,graphicx]{revtex4-2}

\usepackage{graphicx}
\usepackage{dcolumn}
\usepackage{bm}
\usepackage{hyperref}
\usepackage{amsmath}
\usepackage{mathrsfs}
\usepackage[cmintegrals]{newtxmath}
\usepackage[mathlines]{lineno}

\begin{document}

\bibliographystyle{unsrt}
\preprint{APS/123-QED}

\title{Enhanced sensing mechanism based on shifting an exceptional point}

\author{Xuan Mao}
\email{These authors contributed equally to this work}
\affiliation{Department of Physics, State Key Laboratory of Low-Dimensional Quantum Physics, Tsinghua University, Beijing 100084, China}

\author{Guo-Qing Qin}
\email{These authors contributed equally to this work}
\affiliation{Beijing Institute of Radio Measurement, The second Academy of China Aerospace Science and Industry Corporation (CASIC), Beijing 100854, China}

\author{Hao Zhang}
\affiliation{Purple Mountain Laboratories, Nanjing 211111, China}

\author{Bo-Yang Wang}
\affiliation{Department of Physics, State Key Laboratory of Low-Dimensional Quantum Physics, Tsinghua University, Beijing 100084, China}

\author{Dan Long}
\affiliation{Department of Physics, State Key Laboratory of Low-Dimensional Quantum Physics, Tsinghua University, Beijing 100084, China}

\author{Gui-Qin Li}
\affiliation{Department of Physics, State Key Laboratory of Low-Dimensional Quantum Physics, Tsinghua University, Beijing 100084, China}
\affiliation{Frontier Science Center for Quantum Information, Beijing 100084, China}


\author{Gui-Lu Long}
\email{gllong@tsinghua.edu.cn}
\affiliation{Department of Physics, State Key Laboratory of Low-Dimensional Quantum Physics, Tsinghua University, Beijing 100084, China}
\affiliation{Frontier Science Center for Quantum Information, Beijing 100084, China}
\affiliation{Beijing Academy of Quantum Information Sciences, Beijing 100193, China}

\date{\today}

\begin{abstract}

  Non-Hermitian systems associated with exceptional points (EPs) are expected to demonstrate a giant response enhancement for various sensors. The widely investigated enhancement mechanism based on diverging from an EP should destroy the EP and further limits its applications for multiple sensing scenarios in a time sequence. To break the above limit, here we proposed a new enhanced sensing mechanism based on shifting an EP. Different from the mechanism of diverging from an EP, our scheme is an EP non-demolition and the giant enhancement of response is acquired by a slight shift of the EP along the parameter axis induced by perturbation. The new sensing mechanism can promise the most effective response enhancement for all sensors in the case of multiple sensing in a time sequence. To verify our sensing mechanism, we construct a mass sensor and a gyroscope with concrete physical implementations. Our work will deepen the understanding of EP-based sensing and inspire designing various high sensitivity sensors in different physical systems.

\end{abstract}

\maketitle

\section{introduction}

Non-Hermitian quantum systems present attractive properties such as exceptional points (EPs) \cite{miri2019exceptional, wang2021coherent,lai2019observation, chen2017exceptional, ozdemir2019parity, peng2014parity} at which two or more eigenvalues and the corresponding eigenstates coalesce simultaneously \cite{berry2004physics}. The existence of EPs has been verified in various physical systems such as optomechanical systems \cite{xu2016topological, zhang2022dissipative, lu2015p}, coupled atom-cavity systems \cite{choi2010quasieigenstate}, and plasmonic systems \cite{park2020symmetry, moritake2023switchable, li2019exceptional}. The exotic properties of EPs open potential possibilities for advanced applications including phonon lasers \cite{zhang2018phonon, zhang2022dissipative, lu2017exceptional, wang2018polarization}, chiral mode conversion \cite{song2021plasmonic, xu2016topological, zhang2019dynamically, liu2020efficient, hassan2017dynamically, long2022dynamical}, and enhanced sensing \cite{chen2017exceptional, lai2019observation, hodaei2017enhanced, xiao2019enhanced, wiersig2014enhancing, qin2021experimental, djorwe2019exceptional, qin2019brillouin}. The response enhancement at EPs has been widely investigated in microcavities \cite{chen2017exceptional, wiersig2014enhancing, lai2019observation, ren2017ultrasensitive}, optomechanical systems \cite{djorwe2019exceptional, kononchuk2020orientation, mao2020enhanced}, and circuits \cite{xiao2019enhanced, li2023stochastic, nikzamir2021demonstration, bai2023nonlinearity}.

Owing to the complex response near an EP, any perturbation destroys the degeneracy which results in a frequency splitting of the order of the square root of the perturbation strength \cite{chen2017exceptional}. For sufficiently small perturbation, the splitting is much larger than observed in conventional schemes where the response is proportional to the perturbation strength. The giant enhancement of EPs has been proven both theoretically and experimentally in various detection schemes ranging from mass sensor \cite{djorwe2019exceptional} to nanoparticle detector \cite{chen2017exceptional, wiersig2014enhancing, qin2019brillouin}, from magnetometer \cite{kononchuk2022exceptional, kononchuk2020orientation} to gyroscope \cite{lai2019observation, hokmabadi2019non, wang2020petermann, ren2017ultrasensitive, mao2020enhanced, mao2022experimental}. The presence of perturbation will not only cause the frequency splitting, the degeneracy of the linewidths also disappears after introducing perturbation. The EP condition is totally destroyed by the perturbation and the EP no longer exists which will limit the enhancement for multiple sensing schemes in a time sequence. In this case, it is necessary to develop an EP condition nondemolition enhanced sensing mechanism to break the limit.

In this paper, we propose a novel EP-based enhanced sensing mechanism that relies on the shift of an EP in contrast to the conventional method of diverging from an EP. The perturbation breaks the EP condition for diverging from an EP mechanism which deteriorates the EP enhance performance for implementing multiple sensing in a time sequence. Compared with diverging from an EP mechanism, the enhanced sensing mechanism based on shifting an EP demonstrates a slight shift along the parameter axis induced by the perturbation resulting in a remarkable enhancement for the frequency splitting. For the shift of an EP, there exists an EP in parameter space after perturbation and one can prepare the system back to an EP state which promises the most effective enhancement for every sensing. Furthermore, we propose a mass sensor and a gyroscope scheme based on shifting an EP to verify the enhanced sensing mechanism. This work opens up paths to design ultra-sensitive sensors and verifies potential applications in various fields including precision measurement and quantum metrology.

This paper is organized as follows: In Sec. \ref{basic model}, we reveal two enhanced sensing mechanisms based on an EP. We discuss mass sensor and gyroscope scheme based on shifting an EP in Sec. \ref{mass sensor} and Sec. \ref{gyroscope sensor}, respectively. The conclusion is presented in Sec. \ref{conclusion}.

\section{two enhanced sensing mechanisms based on an EP\label{basic model}}

\begin{figure}
  \centering
  \includegraphics[width=\linewidth]{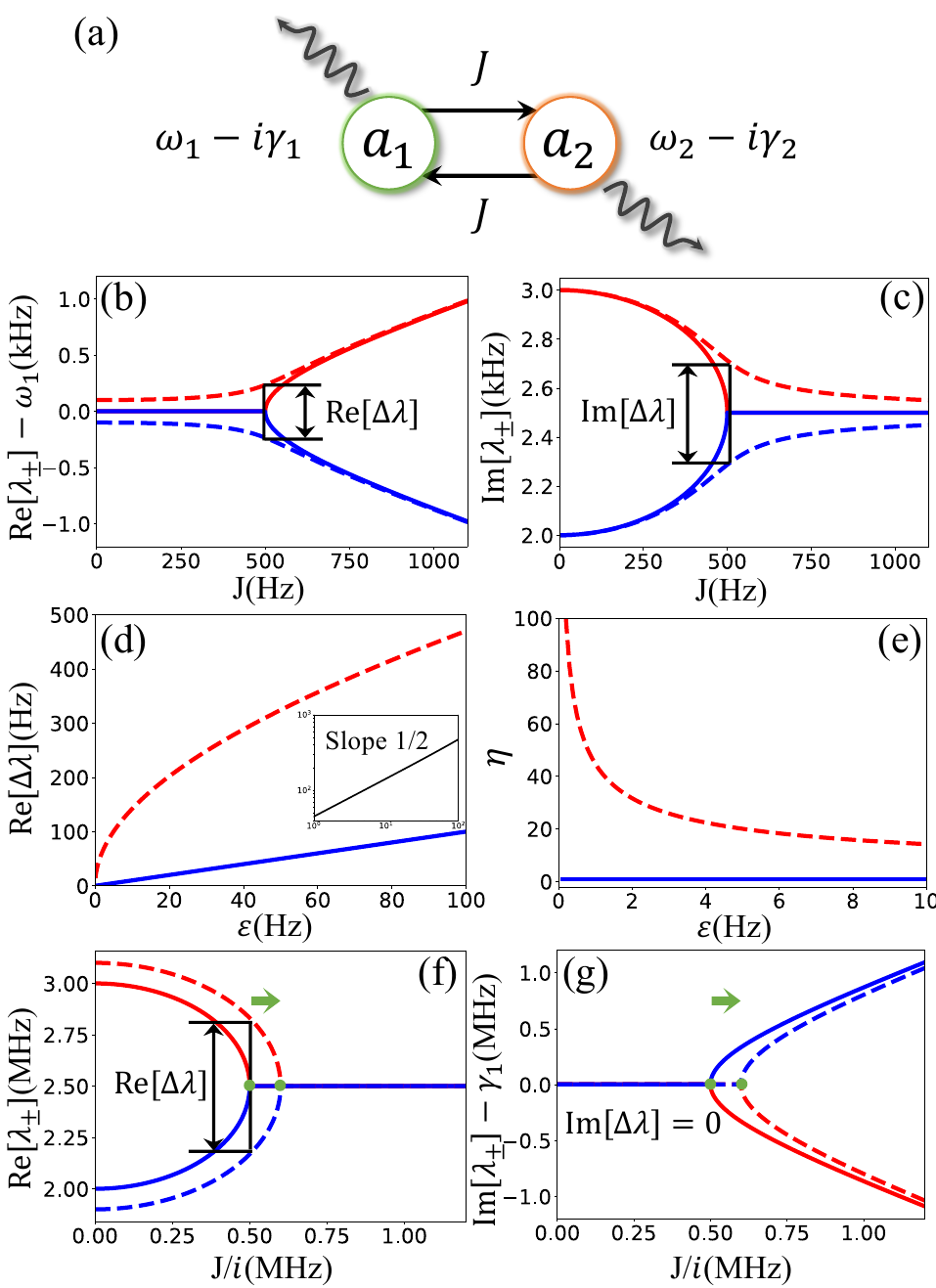}
  \caption{(a) The schematic of the coupled two-level system. The real parts (b) and the imaginary parts (c) of the eigenvalues versus the coupling strength $J$ before (solid lines) and after (dashed lines) the perturbation for the enhanced sensing mechanism based on diverging from an EP. $\varepsilon = 100$ Hz. (d) Illustration of $\mathrm{Re}[\Delta \lambda]$ versus the strength of the perturbation $\varepsilon$ with (red dashed line) and without (blue solid line) an EP enhancement. The inset indicates $\mathrm{Re}[\Delta \lambda]$ versus $\varepsilon$ in a logarithmic scale. (e) The enhancement of sensors based on an EP (red dashed line) and a DP (blue solid line). $J = 500$ Hz. The other parameters are $\omega_1 = \omega_2 = 3$ MHz, $\gamma_1 = 3$ kHz, $\gamma_2 = 2$ kHz. The real parts (f) and the imaginary parts (g) of the eigenvalues versus the coupling strength $J$ before (solid lines) and after (dashed lines) the perturbation for the enhanced sensing mechanism based on shifting an EP. The green dots represent the location of EPs. The parameters are $\omega_1 = 3$ MHz, $\omega_2 = 2$ MHz, $\gamma_1 = \gamma_2 = 3$ kHz, and $\varepsilon = 100$ kHz.}
  \label{model}
\end{figure}

Considering a two-level coupled system composed of two modes represented by $a_1$ and $a_2$ as illustrated by the inset of Fig. \ref{model}(a). The dynamical equations of the general coupled system can be written as
\begin{align}
  \frac{d}{d t}
  \begin{pmatrix}
    a_1 \\
    a_2
  \end{pmatrix}
  = - i
  \begin{pmatrix}
    \omega_1 - i \gamma_1 & J                     \\
    J                     & \omega_2 - i \gamma_2
  \end{pmatrix}
  \begin{pmatrix}
    a_1 \\
    a_2
  \end{pmatrix},
\end{align}
where $t$ can be the evolution time and the propagation distance. The resonance frequency and the decay rate of $a_1$ ($a_2$) are represented by $\omega_1$ ($\omega_2$) and $\gamma_1$ ($\gamma_2$), respectively. $J$ denotes the coupling coefficient and it can be real or complex. For the purpose of concise, we only consider the reciprocal coupling case and the similar conclusion can be obtained for the nonreciprocal coupling. Assuming the solutions have the harmonic form $a_{1,2} = a_{1,2} e^{-i \lambda t}$ and the eigenvalues of the coupled system can be expressed as
\begin{align}
  \lambda_{\pm} = \omega_{\mathrm{mean}} - i \gamma_{\mathrm{mean}} \pm \sqrt{(\omega_{\mathrm{diff}} + i \gamma_{\mathrm{diff}})^2 + J^2}.
  \label{eigenvalues}
\end{align}
$\omega_{\mathrm{mean}} = (\omega_1 + \omega_2) / 2$ and $\gamma_{\mathrm{mean}} = (\gamma_1 + \gamma_2) / 2$ represent the mean values of the frequency and the loss rate. $\omega_{\mathrm{diff}} = (\omega_1 - \omega_2) / 2$ and $\gamma_{\mathrm{diff}} = (\gamma_1 - \gamma_2) / 2$ are the differences of the resonance frequency and the decay rate. It can be indicated that when the parameter values meet the requirement $\pm i J = \omega_{\mathrm{diff}} + i \gamma_{\mathrm{diff}}$ the system is prepared on an EP state, at which the real parts and the imaginary parts of the eigenvalues coalesce simultaneously.

If the coupling coefficient $J$ is a real number and the enhanced sensing mechanism is well-known as the eigenvalues diverging from an EP. The enhanced sensing mechanism based on the eigenvalues diverging from an EP after the perturbation $\varepsilon$ is illustrated by Figures \ref{model}(b) and \ref{model}(c). The solid and the dashed lines represent the evolution of the two eigenvalues along with the system parameter $J$ before and after the perturbation. Usually, before introducing perturbation, the sensing system is prepared near an EP and one can monitor the frequency shift $\mathrm{Re}[\Delta \lambda]$ and the linewidth change $\mathrm{Im}[\Delta \lambda]$ after the perturbation. According to the frequency shift, the strength of the perturbation $\varepsilon$ can be indicated. Due to the singularity of EPs, the frequency shifts of the eigenvalues obtain $ \varepsilon^{1/2}$ (the red line in Fig. \ref{model}(d)) enhancement comparing to the diabolic point (DP) (the blue solid line in Fig. \ref{model}(d)) in the condition of $ \varepsilon \ll 1$. The inset of Fig. \ref{model}(d) indicates the frequency shift is proportional to $\varepsilon^{1/2}$ as the slope is $1/2$ in a logarithmic scale. To demonstrate the enhancement of EPs, we can define the enhancement factor
\begin{align}
  \eta = |\frac{\mathrm{Re}[\Delta \lambda]}{\varepsilon}|.
\end{align}
Fig. \ref{model}(e) illustrates the enhancement factor of an EP (the red dashed line) and a DP (the blue solid line) versus the perturbation strength. It can be indicated an EP enhance performance is much better in the weak perturbation regime which is consistent with the analysis above. It can be inferred that the EP condition is destroyed by the perturbation and the EP no longer exists in the parameter space for the enhanced sensing mechanism based on diverging from an EP. In the case of implementing multiple sensing in a time sequence, the enhancing performance deteriorates as the number of sensing increases as every next detection is conducted on the basis of being destroyed by the previous detection.

The shift of an EP mechanism requires the coupling coefficient is an imaginary number and can also exhibit $\varepsilon^{1/2}$ enhancement. Under some specific parameter values, the existence of the EP can be maintained along with a shift in the parameter space after the perturbation which is denoted by Figures \ref{model}(f) and \ref{model}(g). Due to the topological structure near an EP, the shift of an EP will lead to similar enhancement results to the sensing performance based on the eigenvalues diverge. On the other hand, the linewidth changes of the two eigenstates demonstrate different behavior. In the case of an EP shift sensing, the two linewidths of the eigenstates stay the same as before the perturbation which may further improve the enhancement performance in the perspective of experiments. The unbroaden linewidthes after the perturbation may improve the precision of dispersion measurement. Another interesting point is there also exists an EP in the parameter space after the perturbation for the sensors based on shifting an EP which is an EP condition nondemolition enhanced sensing mechanism. One can prepare the system back to an EP state by tuning corresponding parameter after introducing the perturbation. In the case of the influence of the perturbation to the sensor is hard to eliminate such as mass sensor and nanoparticle sensing, it is a challenge task to separate the deposited mass or nanoparticle and the sensing system, the sensors based on shifting an EP will always obtain the most efficient enhancement as the system can be prepared back to an EP before the next sensing. Thus, for the task of implementing multiple sensing in a time sequence, sensors based on shifting an EP will always demonstrate the most effective EP enhancement for every detection.

\section{mass sensor based on shifting an EP\label{mass sensor}}

\begin{figure*}
  \centering
  \includegraphics[width=\linewidth]{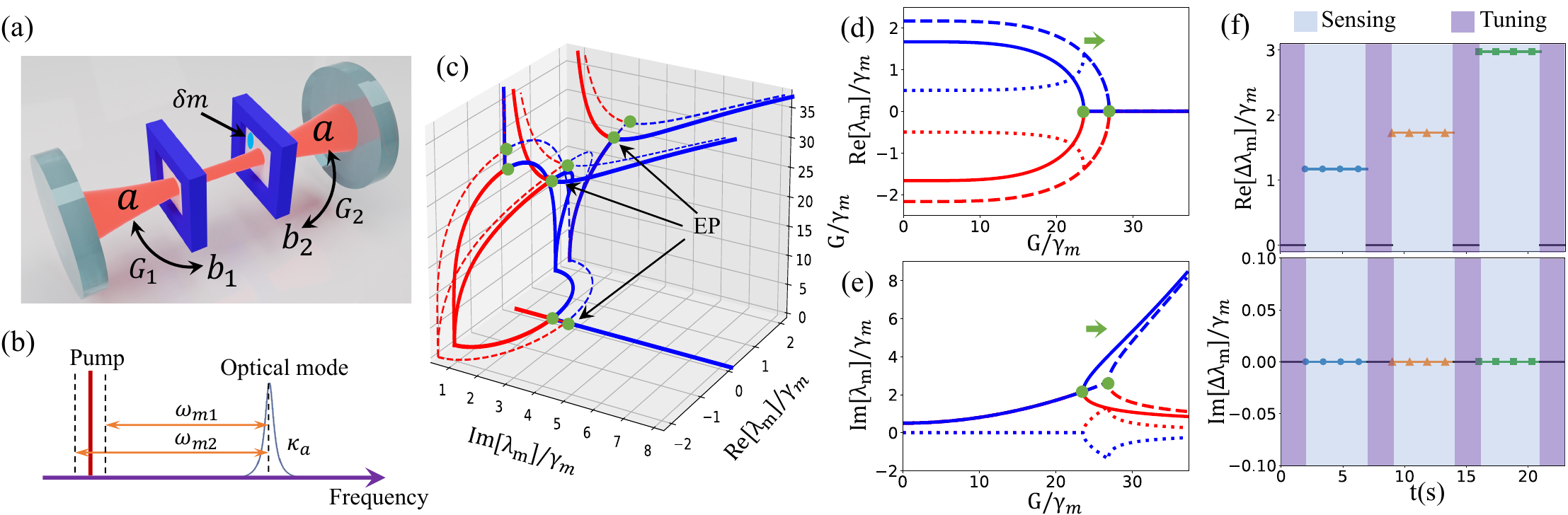}
  \caption{(a) The mass sensor schematic. (b) The pump scheme in frequency domain. (c) The evolution of the eigenvalues under different optomechanical coupling strength $G$. EPs are denoted by the green dots. (d) and (e) illustrate the real parts and the imaginary parts of the mass sensor system versus $G$. The solid lines and the dashed lines represent before and after introducing a deposited mass, respectively. The dotted lines are the differences between the eigenvalues before and after the perturbation. $\delta \omega = 0.0005 \omega_1$. (f) demonstrates the real parts and the imaginary parts difference for implementing multiple mass sensing in a time sequence. The corresponding frequency shifts are $\delta \omega = 0.0001 \omega_1, 0.0002 \omega_1$ and $0.0005 \omega_1$, respectively. The other parameters are $\omega_{m1} / 2 \pi = 6$ GHz, $\omega_{m2} / 2 \pi = 6.01$ GHz, $\gamma_{m1} / 2 \pi = \gamma_{m2} / 2 \pi = \gamma_m / 2 \pi = 3$ MHz, $\kappa_a / 2 \pi = 2$ GHz.}
  \label{mass}
\end{figure*}

Optomechanics \cite{aspelmeyer2014cavity} has been a promising platform to investigate fundamental phenomena \cite{weis2010optomechanically, kim2015non, kronwald2013optomechanically, qin2020manipulation} and applied science \cite{del2007optical, wang2012using, tian2012adiabatic, ruesink2018optical, shen2018reconfigurable, domeneguetti2021parametric}. Benefiting from abundant manipulation methods, realization of ultra-sensitive sensors operating at EPs in optomechanical systems has gained wide attention over last few decades. To illustrate the enhanced mechanism based on shifting an EP, here we propose an optomechanical mass sensor. The mass sensor system as illustrated by Fig. \ref{mass}(a) consists of two mechanical vibrators $b_1$ and $b_2$ coupling to the same optical field $a$ with frequency $\omega_a$ and loss rate $\kappa_a$. The Hamiltonian of the system is ($\hbar = 1$)
\begin{align}
  H_m = & \sum_{j = 1, 2} \big[\omega_{mj} b^{\dagger}_j b_j + g_j a^{\dagger} a (b_j + b^{\dagger}_j)\big]\nonumber \\ &+ \omega_a a^{\dagger} a + i \sqrt{\kappa_{\mathrm{ex}}} \epsilon (a^{\dagger} e^{-i \omega_d t} - a e^{i \omega_d t}).
\end{align}
The mechanical resonator $b_j$ characterize by the frequency $\omega_{mj}$ and the damping rate $\gamma_{mj}$ ($j = 1, 2$). $g_j$ is the single-photon optomechanical coupling strength between the mechanical mode $b_j$ and the optical mode $a$. $\kappa_{\mathrm{ex}}$ represents the coupling loss. Fig. \ref{mass}(b) represents the driving scheme of the mass sensor in frequency domain. The pump laser owns the driving frequency $\omega_d$ and the amplitude $\epsilon$. Following the standard optomechanical linearization procedure and adiabatically eliminate the optical field, we can obtain the effective Hamiltonian of the two mechanical resonators in matrix form
\begin{align}
  H^{\mathrm{eff}}_m =
  \begin{pmatrix}
    \Omega_0 - i \bigg(\frac{\gamma_{m1}}{2} + \frac{2 G^2_1}{\kappa_a} \bigg) & - i \frac{2 G_1 G_2}{\kappa_a}                                               \\
    - i \frac{2 G_1 G_2}{\kappa_a}                                             & - \Omega_0 - i \bigg(\frac{\gamma_{m2}}{2} + \frac{2 G^2_2}{\kappa_a} \bigg)
  \end{pmatrix}\label{meff},
\end{align}
where $\Omega_0 = (\omega_{m1} - \omega_{m2})/2$. $G_1$ and $G_2$ are the effective optomechanical coupling strengths. For convenience, we can set $G_1 = G_1 = G$ and $\gamma_{m1} = \gamma_{m2} = \gamma_m$. The eigenvalues of $H^{\mathrm{eff}}_m$ are given by
\begin{align}
  \lambda_{m\pm} = - i (\gamma_m/2 + \Gamma) \pm \sqrt{\Omega_0^2 - \Gamma^2},
\end{align}
where $\Gamma = 2 G^2 / \kappa_a$. Note the Hamiltonian form in Eq. \ref{meff} satisfies anti-PT symmetry with an EP at $\Gamma = |\Omega_0|$. The evolution of the eigenvalues under different optomechanical coupling strength are illustrated by the solid lines in Figures \ref{mass}(c)-\ref{mass}(e). Note EPs of the system are marked by the green dots in Fig. \ref{mass}(c).

For a traditional mass sensor, the relationship between the deposited mass $\delta m$ and the caused frequency shift $\delta \omega$ can be given by \cite{li2007ultra}
\begin{align}
  \delta \omega = \frac{\omega_{m2}}{2 m} \delta m = R \delta m .
\end{align}
$R = \omega_{m2} / 2 m$ represents the mass responsivity. $\omega_2$ and $m$ are the resonate frequency and the mass of the mechanical resonator supporting the deposition.

After depositing mass $\delta m$ into the mechanical resonator $b_2$, the EP will move from the former point in the parameter space as Figures \ref{mass}(c)-\ref{mass}(e) shown. The dashed lines demonstrate the evolution of the eigenvalues after the perturbation. In Figures \ref{mass}(d) and \ref{mass}(e), the dotted lines show the difference between the eigenvalues before and after, i. e. $\mathrm{Re}[\lambda_m]_{\mathrm{after}} - \mathrm{Re}[\lambda_m]_{\mathrm{before}}$ and $\mathrm{Im}[\lambda_m]_{\mathrm{after}} - \mathrm{Im}[\lambda_m]_{\mathrm{before}}$. It can be noticed that the deposited mass causes the shift of the EP and further lead to the largest frequency shift near the EP. Meanwhile, the introducing of the deposited mass will not change the bandwidth of the eigenstate, which theoretically boosts the sensitivity enhancement for EP-based sensors.

One can implement multiple mass sensing in a time sequence utilizing the mass sensor based on shifting an EP as it can be tuned back to an EP after each sensing period. Fig. \ref{mass}(f) demonstrates the real parts and the imaginary parts difference. There are two periods, sensing period and tuning period, during the time sequence mass sensing. During the sensing periods the frequency shifts obtain the most effective enhancement and present mass dependent response. The corresponding frequency shifts due to the deposited mass are $\delta \omega = 0.0001 \omega_1, 0.0002 \omega_1$ and $0.0005 \omega_1$, respectively. During the tuning periods the system can be prepared back to an EP state by tuning the optomechanical coupling strength. Note that no matter the system experiences sensing periods or tuning periods, the linewidths of the sensor are always the same as the unperturbed system as shown in Fig. \ref{mass}(f).

\section{gyroscope based on shifting an EP\label{gyroscope sensor}}

\begin{figure*}
  \centering
  \includegraphics[width=\linewidth]{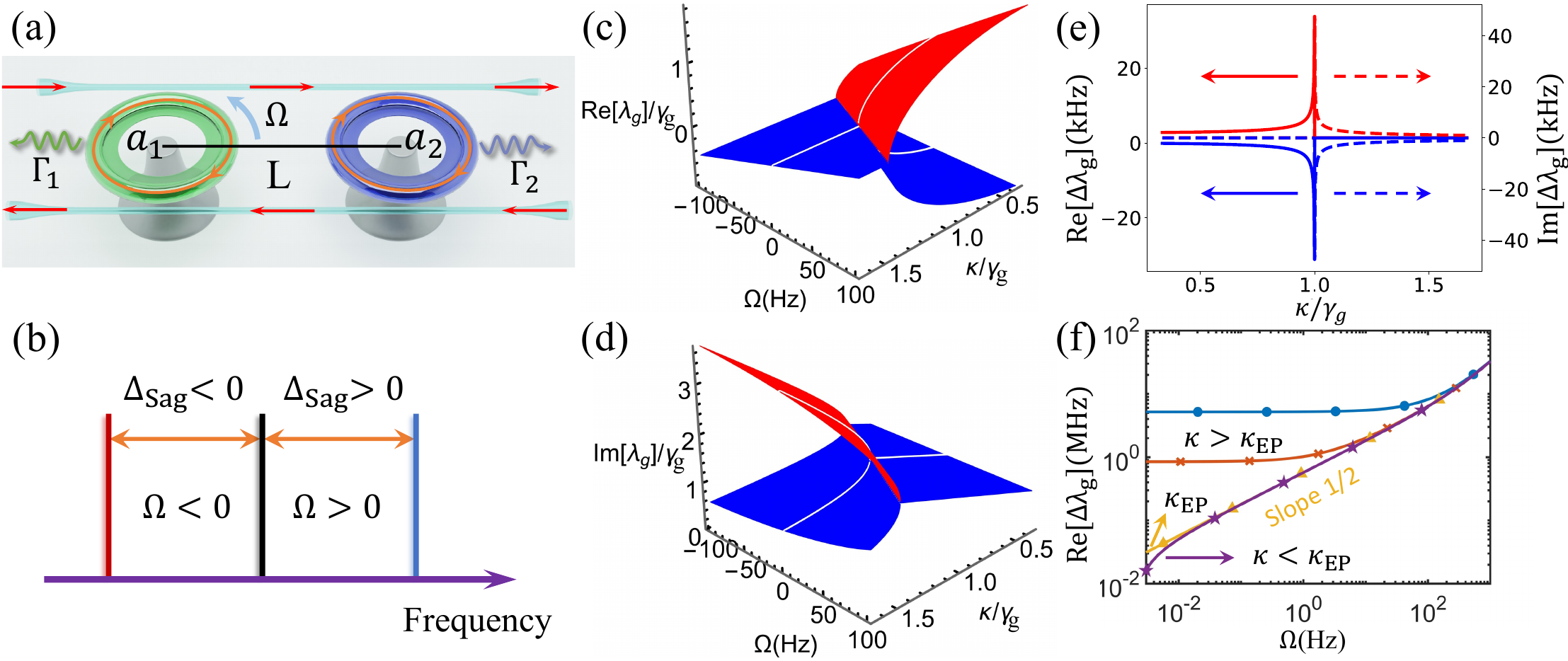}
  \caption{(a) The gyroscope schematic. (b)The Sagnac frequency shift due to the introduction of rotation. (c) and (d) The real parts and the imaginary parts of the eigenvalues of the system. (e) The differences of the real parts and the imaginary parts of the eigenvalues before and after rotation under different values of $\kappa$. $\Omega = 0.1$ Hz. (f) Dependance of frequency splitting on the rotation frequency $\Omega$ in the case of anti-PT broken phase ($\kappa < \kappa_{\mathrm{EP}}$), anti-PT symmetry phase ($\kappa > \kappa_{\mathrm{EP}}$), and at an EP ($\kappa_{\mathrm{EP}}$). The other parameters are $n = 1.44$, $R = 9$ mm, $c = 3 \times 10^8$ m/s, $\lambda = 1550$ nm, and $\Omega_- = \Gamma_1 = \Gamma_2 = \gamma_g = 3$ MHz.}
  \label{gyroscope}
\end{figure*}

The gyroscope scheme based on shifting an EP is illustrated by Fig. \ref{gyroscope}(a). Two WGM cavities are evanescently coupled to two common drop-filter waveguides with center to center distance $L$. In the basis of two resonator modes $\Psi = (a_1, a_2)^{\top}$, the effective Hamiltonian in matrix form can be expressed as \cite{peng2020level}
\begin{align}
  H^{\mathrm{eff}}_g =
  \begin{pmatrix}
    \omega_{a1} - i \frac{\gamma_{\mathrm{eff1}}}{2} & - i e^{i\varphi} \sqrt{\kappa_1 \kappa_2}        \\
    - i e^{i\varphi} \sqrt{\kappa_1 \kappa_2}        & \omega_{a2} - i \frac{\gamma_{\mathrm{eff2}}}{2}
  \end{pmatrix},
\end{align}
where $\omega_{a1}$ and $\omega_{a2}$ are the resonance frequency of the two WGM resonators. $\gamma_{\mathrm{effj}} = \Gamma_j + 2 \kappa_j$ denotes the effective loss and $\Gamma_j$ represents the intrinsic loss rate in the $j$-th resonator. $\kappa_j$ is the coupling rate between the $j$-th resonator and the drop-filter waveguide. The phase delay factor is $\varphi = 2 \pi n_w L / \lambda$, with $n_w$ being the refractive index of waveguides and $\lambda$ represents the wavelength of the light.

Benefiting from the adjustable of the phase delay factor, the indirectly coupling system may exhibit anti-PT symmetry \cite{peng2016anti, li2019anti, zhang2022dissipative, wu2023chip}. Without
losing of generality, we can assume the two resonators have the equal losses $\Gamma_1 = \Gamma_2$ and the waveguide-resonator coupling rates are equal $\kappa_1 = \kappa_2 = \kappa$. Thereby the effective Hamiltonian in anti-PT symmetry form is
\begin{align}
  H^{\mathrm{eff}^\prime}_g =
  \begin{pmatrix}
    \Omega_{-} - i \bigg(\frac{\Gamma_1 + \Gamma_2}{4} + \kappa \bigg) & i \kappa                                                            \\
    i \kappa                                                           & -\Omega_{-} - i \bigg(\frac{\Gamma_1 + \Gamma_2}{4} + \kappa \bigg)
  \end{pmatrix}\label{geff}.
\end{align}
Here $\Omega_{-} = (\omega_{a1} - \omega_{a2}) / 2$. When introducing rotation frequency into the first WGM resonator, the resonance frequency experiences a Sagnac frequency shift associated with the rotation direction
\begin{align}
  \Delta_{\mathrm{Sag}} = \frac{n R \Omega \omega_{a1}}{c} \bigg(1 - \frac{1}{n^2} - \frac{\lambda}{n} \frac{\mathrm{d}n}{\mathrm{d}\lambda} \bigg),
\end{align}
where $n$ and $R$ are the refractive index and radius of the resonator, respectively. $c$ denotes the speed of the light and the dispersion term $\mathrm{d}n/\mathrm{d}\lambda$ is relatively small in typical materials ($\sim 1 \%$) \cite{maayani2018flying}. For convenience, we define the sign of the rotation frequency depending on the rotation direction. As shown in Figures \ref{gyroscope}(a) and \ref{gyroscope}(b), in the case of counterclockwise rotation the sign of the rotation frequency is defined as positive associating with a positive Sagnac frequency shift and vice versa. Due to the definition of the sign of the rotation frequency, one can simply add the perturbation matrix into Eq. \ref{geff} after introducing rotation
\begin{align}
  H^{\mathrm{eff}^{\prime \prime}}_g =
  \begin{pmatrix}
    \Omega_{-} +\Delta_{\mathrm{Sag}} - i \bigg(\frac{\Gamma_1 + \Gamma_2}{4} + \kappa \bigg) & i \kappa                                                            \\
    i \kappa                                                                                  & -\Omega_{-} - i \bigg(\frac{\Gamma_1 + \Gamma_2}{4} + \kappa \bigg)
  \end{pmatrix}.
\end{align}
And the eigenvalues of the effective Hamiltonian are
\begin{align}
  \lambda_{g \pm} = \frac{\Delta_{\mathrm{Sag}}}{2} - i \bigg(\frac{\Gamma_1 + \Gamma_2}{4} + \kappa \bigg) \pm \sqrt{\bigg(\frac{\Delta_{\mathrm{Sag}} + 2 \Omega_{-}}{2} \bigg)^2 - \kappa^2}.
\end{align}

It is clear that no matter the system experiences rotation or not there is always a threshold coupling strength at $(\Delta_{\mathrm{Sag}} + 2\Omega_{-})/2$. The real parts and the imaginary parts of $\lambda_{g \pm}$ are illustrated by Figures \ref{gyroscope}(c) and \ref{gyroscope}(d). In the absence of the rotation, there are three regimes for different values of $\kappa$ and the threshold is $\kappa_{\mathrm{EP}} = \Omega_{-}$.  In the domain of $\kappa < \kappa_{\mathrm{EP}}$ the eigenstates share the same linewidth while possess different resonance frequency and the system is in the anti-PT broken phase. In the case of $\kappa > \kappa_{\mathrm{EP}}$ the two eigenfrequencies have equal real part and different imaginary parts. The EP is the phase transition point of anti-PT symmetry phase and anti-PT broken phase. The location of the EP is moving in the parameter space as the rotation frequency varies. Fig. \ref{gyroscope}(e) shows the differences of the real parts and the imaginary parts of the eigenvalues before and after rotation, i. e. $\mathrm{Re}[\lambda_g]_{\mathrm{after}} - \mathrm{Re}[\lambda_g]_{\mathrm{before}}$ and $\mathrm{Im}[\lambda_g]_{\mathrm{after}} - \mathrm{Im}[\lambda_g]_{\mathrm{before}}$. The solid lines and the dashed lines are corresponding to the real parts and the imaginary parts, respectively. One can see near an EP the system demonstrates a great frequency splitting and maintains the linewidth meanwhile.

After introducing rotation into the system, the reactions of the frequency splitting $\mathrm{Re}[\Delta \lambda] = |\mathrm{Re}[\lambda_{g+}] - \mathrm{Re}[\lambda_{g-}]|$ in three regimes are not the same. Fig. \ref{gyroscope}(f) illustrates the logarithmic behavior of the frequency splitting under different rotation in the case of anti-PT symmetry phase ($\kappa > \kappa_{\mathrm{EP}}$), anti-PT broken phase ($\kappa < \kappa_{\mathrm{EP}}$), and at the EP ($\kappa_{\mathrm{EP}}$). In the case of anti-PT symmetry phase, the frequency splitting is largest while the slope is smallest for the same perturbation strengths. However, for a small rotation frequency the slope of response at an EP is $1/2$ which can be explained by the perturbation theory \cite{chen2017exceptional}. For a relatively large rotation frequency, the perturbation theory no longer works and therefore the slope of the response is larger than $1/2$.

\section{conclusion\label{conclusion}}

We propose a novel enhanced sensing mechanism based on perturbation induced shift of an EP in contrast to the widely investigated method of diverging from an EP. Diverging from an EP mechanism destroys the EP condition in the parameter space and obtains $\varepsilon^{1/2}$ enhancement in the weak perturbation regime. In this case, the EP condition is totally destroyed by the perturbation and the EP no longer exists which will limit the enhancement for multiple sensing schemes in a time sequence. To overcome the challenge, we propose the EP condition nondemolition enhanced sensing mechanism based on shifting an EP which demonstrates a slight shift along the parameter axis induced by the perturbation and also exhibits remarkable enhancement. The linewidthes will maintain after the perturbation for the EP shift mechanism, which may improve the precision of dispersion measurement. To implement the EP shift sensing mechanism, we construct a mass sensor and a gyroscope sensing scheme for the mechanical modes and optical mode pair, respectively. Combining other enhancing methods such as exceptional surfaces \cite{qin2021experimental, zhong2019sensing, zhang2019experimental} and nonlinearity \cite{zhang2020breaking, silver2021nonlinear, bai2023nonlinearity}, the performance of sensors will be further improved. The enhanced sensing mechanism proposed in this paper provides a comprehensive understanding of EPs enhancing and may inspire technological development in various schemes.

\begin{acknowledgments}

  This work is supported by the National Natural Science Foundation of China (6213100); The Key  Research and Development Program of Guangdong province (2018B030325002); Beijing Advanced Innovation Center for Future Chip (ICFC); Tsinghua University Initiative Scientific Research Program.

\end{acknowledgments}

\nocite{*}



%

\end{document}